\documentclass[journal=nalefd,manuscript=letter]{achemso}
\usepackage{amsmath}
\usepackage[version=3]{mhchem} % Formula subscripts using \ce{}
\usepackage[squaren]{SIunits}
\usepackage{textgreek}
\usepackage{dcolumn}
\usepackage{bm}
\usepackage{braket}

\author{Andreas W. Schell}
\email{andreas.schell@physik.hu-berlin.de}
\affiliation{Department of Electronic Science and Engineering, Kyoto University, Kyoto daigaku-katsura, Nishikyo-ku, Kyoto, Japan}
\alsoaffiliation{Research Institute for Electronic Science, Hokkaido University, Sapporo, Hokkaido, Japan}
\alsoaffiliation{The Institute of Scientific and Industrial Research,
Osaka University, Mihogaoka 8-1, Ibaraki, Osaka, Japan}
\altaffiliation{These authors contributed equally.}

\author{Hideaki Takashima}
\affiliation{Department of Electronic Science and Engineering, Kyoto University, Kyoto daigaku-katsura, Nishikyo-ku, Kyoto, Japan}
\alsoaffiliation{Research Institute for Electronic Science, Hokkaido University, Sapporo, Hokkaido, Japan}
\alsoaffiliation{The Institute of Scientific and Industrial Research,
Osaka University, Mihogaoka 8-1, Ibaraki, Osaka, Japan}
\altaffiliation{These authors contributed equally.}

\author{Shunya Kamioka}
\affiliation{Research Institute for Electronic Science, Hokkaido University, Sapporo, Hokkaido, Japan}
\alsoaffiliation{The Institute of Scientific and Industrial Research,
Osaka University, Mihogaoka 8-1, Ibaraki, Osaka, Japan}
\altaffiliation{These authors contributed equally.}

\author{Yasuko Oe}
\affiliation{Department of Electronic Science and Engineering, Kyoto University, Kyoto daigaku-katsura, Nishikyo-ku, Kyoto, Japan}
\alsoaffiliation{Research Institute for Electronic Science, Hokkaido University, Sapporo, Hokkaido, Japan}
\alsoaffiliation{The Institute of Scientific and Industrial Research,
Osaka University, Mihogaoka 8-1, Ibaraki, Osaka, Japan}

\author{Masazumi Fujiwara}
\affiliation{Nano-Optics, Institute of Physics, Humboldt-Universit\"{a}t zu
Berlin, Newtonstra{\ss}e~15, Berlin, Germany}
\alsoaffiliation{Research Institute for Electronic Science, Hokkaido University, Sapporo, Hokkaido, Japan}
\alsoaffiliation{The Institute of Scientific and Industrial Research,
Osaka University, Mihogaoka 8-1, Ibaraki, Osaka, Japan}

\author{Oliver Benson}
\affiliation{Nano-Optics, Institute of Physics, Humboldt-Universit\"{a}t zu
Berlin, Newtonstra{\ss}e~15, Berlin, Germany}

\author{Shigeki Takeuchi}
\email{takeuchi@kuee.kyoto-u.ac.jp}
\affiliation{Department of Electronic Science and Engineering, Kyoto University, Kyoto daigaku-katsura, Nishikyo-ku, Kyoto, Japan}
\alsoaffiliation{Research Institute for Electronic Science, Hokkaido University, Sapporo, Hokkaido, Japan}
\alsoaffiliation{The Institute of Scientific and Industrial Research,
Osaka University, Mihogaoka 8-1, Ibaraki, Osaka, Japan}

\title{Highly Efficient Coupling of Nanolight Emitters to a Ultra-Wide Tunable Nanofibre Cavity}

\begin{document}

%\textbf{
\begin{abstract}
Solid-state microcavities combining ultra-small mode volume, wide-range
resonance frequency tuning, as well as lossless coupling to a single mode
fibre are integral tools for nanophotonics and quantum networks. 
We developed an integrated system providing all of these three indispensable
properties. It consists of a nanofibre Bragg cavity
(NFBC) with the mode volume of under \unit{1}{\micro\meter^3} and repeatable
tuning capability over more than \unit{20}{\nano\meter} at visible
wavelengths. In order to demonstrate quantum light-matter interaction, we
establish coupling of quantum dots to our tunable NFBC and achieve an
emission enhancement by a factor of 2.7.
\end{abstract}
%\textbf{Keywords} 
%Tapered fibre, microcavities, ultra-small mode volume, resonance frequency tuning, emission enhancement

\section*{Introduction}

Solid-state devices that confine light to a very small volume are 
important to control 
the optical properties of light emitters. Such devices are especially useful  as single photon sources~\cite{Takeuchi2014}, photonic quantum 
memories~\cite{Lvovsky2009}, and photonic quantum gates~\cite{Kok2007} in photonic 
quantum networks~\cite{Kimble2008,O'Brien2009}. 
In those applications, there are three key requirements: an ultra-small mode volume, ideally on the order of 
$\lambda^3$ (where $\lambda$ is the emission wavelength), a wide tuning range of the resonance frequency,
and lossless coupling of photons to a single mode optical fibre (SMF). Here, we report 
on the development of an integrated fibre-coupled all-solid-state device that fulfils  all 
three requirements. First, we fabricate a 
nanofibre Bragg cavity (NFBC)~\cite{LeKien2009,Takeuchi-Patent-2010} 
with a diameter of a few hundred nanometres and a mode volume of less than \unit{1}{\micro\meter^3}. 
Second, we demonstrate a repeatable tuning of its resonance frequency over an ultra-wide range of more than 
\unit{20}{\nano\meter} at a visible wavelength ($\sim$~\unit{640}{\nano\meter}), which is possible even at 
cryogenic temperatures ($\sim$~\unit{85}{\kelvin}). Third, 
we succeeded in efficiently coupling colloidal quantum dots (QDs) to a NFBC and observed emission 
enhancement by a factor of 2.7 due to the NFBC.
NFBCs with broad tuning capability are not only be expected to be key tools for the realization of nanophotonic quantum 
networks~\cite{Kimble2008}, but also for numerous real-world applications in photonics, 
including fibre-embedded zero-threshold microlasers~\cite{Yokoyama1992}, ultra-wide tunable nanofibre-filters, 
and ultra-sensitive nanosensors for life sciences~\cite{Vollmer2012}.
%\end{abstract}

Solid-state microcavities with small mode volume~\cite{Vahala2003} come in different types:  
One- and two-dimensional photonic crystal microcavities~\cite{Riedrich-Moller2012,Akahane2003} 
have recently been used in highly  efficient 
single photon sources~\cite{Malik2010} and all-optical switching by single photons~\cite{Volz2012}. 
However, for these 
devices types, the coupling loss of photons to SMFs, which are crucial 
for most  
applications, especially in nanophotonic quantum networks, remains an important problem to solve. Alternatively, 
microspheres~\cite{Knight1997,Cai2000,Konishi2006} and 
microtoroids~\cite{Armani2003} have high quality factors (Q-factors) and can be coupled to SMFs with small 
coupling loss using tapered optical fibres~\cite{Knight1997}. However, their mode volumes are large 
(typically hundreds times of $\lambda^3$) and precise coupling via tapered 
fibres is a demanding task. Furthermore, the tuning range of all these devices usually is limited 
to about a few nanometres for visible wavelengths and requires  changes of experimental 
conditions such as temperatures~\cite{Park2006,Englund2007} or the surface laminating layer 
of the cavity~\cite{Mosor2005,Srinivasan2007,Faraon2011}. 

Tapered optical fibres are increasingly attracting attention as means to couple photons from light emitters to SMFs. 
When the diameter of the tapered region is about $\lambda/2$, the field intensity of the propagating 
mode becomes strong even outside the fibre. This permits the emission mode from the light emitter on 
 the tapered region's surface to efficiently couple to the propagating mode~\cite{Kien2004}.
Since in this way the light is emitted directly in the mode of the SMF, no
photons are lost due to additional fibre-coupling or 
due to small numerical apertures of the collection optics, as they occur
for example in microscope systems.
 Efficient coupling 
of photons from colloidal QDs~\cite{Fujiwara2011,Yalla2012a,yalla2012efficient} and nitrogen vacancy centres in 
nanodiamonds~\cite{Schroder2012,Liebermeister2014} has also been 
demonstrated. Recently, NFBC devices, where the microcavity is fabricated in the tapered 
region of the tapered fibre has been proposed by us~\cite{Takeuchi-Patent-2010} and 
Hakuta~\cite{LeKien2009} independently, and realization 
of such devices using ion-beam milling has been reported~\cite{K.P.Nayak.FIB2011}.
Very recently, coupling of an emitter to a cavity formed by an external grating for 
modulation the refractive indes has been reported~\cite{Hakuta_PRL}.

In this letter, we report on the ultra-wide and repeatable tuning of an 
NFBC device. We have found that simply by controlling the tension applied to the NFBC, it is possible to 
tune the 
resonance frequency by more than \unit{20}{\nano\meter} at visible wavelengths. 
Furthermore, we have succeeded in coupling colloidal QDs to a tunable NFBC and have 
observed enhanced emission from the hybrid device. 

\section*{Results}

\begin{figure}
	\includegraphics[width=.7 \textwidth,clip]{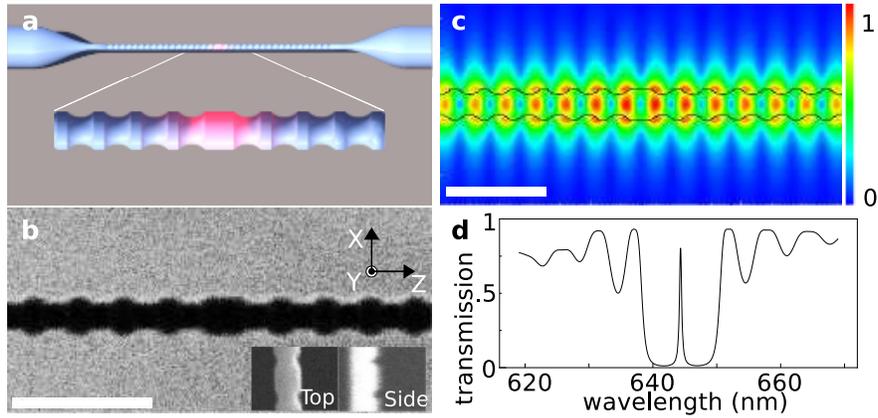}
	\caption{\textbf{Nanofibre Bragg cavities.} 
	\textbf{a}, Sketch of an NFBC. Note that the number of grating periods has been
	reduced.
	\textbf{b}, SIM image of an NFBC 
		(diameter \unit{270}{\nano\meter},  groove depth \unit{45}{\nano\meter},  pitch \unit{300}{\nano\meter}, 
	and defect length \unit{450}{\nano\meter}). 
	Scalebar is \unit{1}{\micro\meter}.  
	\textbf{c},
	distribution
	of the electric field calculated using a 3D finite-difference time-domain (FDTD) algorithm. The black line 
	indicates the cavity structure and the scalebar is 
	\unit{1}{\micro\meter}.  
	\textbf{d},  
	Transmission spectrum of the NFBC as obtained from the FDTD calculations.}
	\label{fig1}
\end{figure}

Our NFBCs are fabricated using a focused ion beam (FIB) milling system 
(see Methods). Typical  nanofibre diameters are around \unit{300}{\nano\meter}, 
which enables single mode operation and efficient coupling to emitters at our target wavelength 
of \unit{630}{\nano\meter}~\cite{Fujiwara2011}. Figure~\ref{fig1}a shows a schematics of an NFBC, and 
Figure~\ref{fig1}b shows a scanning ion microscope (SIM) image of a fabricated device 
(3$\lambda$/4 defect in a 160 period Bragg grating). 
The depth of the groove, the length of the pitch, and the length of the defect are \unit{45}{\nano\meter}, 
\unit{300}{\nano\meter}, and \unit{450}{\nano\meter}, respectively.

\begin{figure}
	\includegraphics[width=.7 \textwidth]{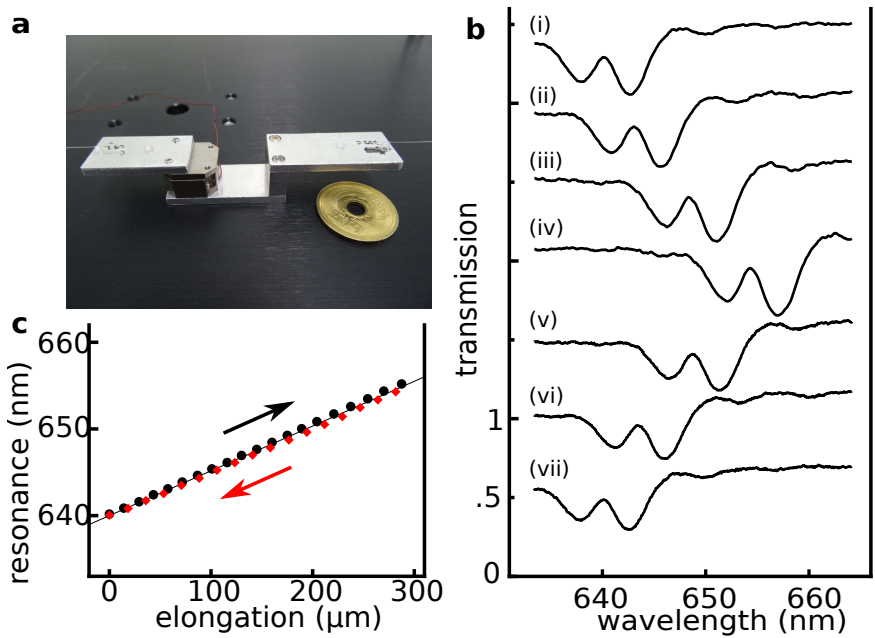}
	\caption{\textbf{Tuning of the fibre cavities.} \textbf{a} Holder used for tensioning the fibre. 
	\textbf{b} Transmission
	spectra  of an NFBC at different tunings. The distances the piezo actuator is moved 
	are \unit{0}{\micro\meter}, \unit{57.5}{\micro\meter}, \unit{160}{\micro\meter}, \unit{270}{\micro\meter}, 
	\unit{176.7}{\micro\meter}, \unit{70.8}{\micro\meter}, and \unit{0}{\micro\meter} for i-vii, respectively.
	\textbf{c} Reversible tuning of an NFBC.
	The behaviour of the resonance peak when the NFBC is tuned is shown.
	Black circles indicate stretching of the fibre while red diamonds indicate compressing.}
	\label{fig2}
\end{figure}

Numerical calculations
indicate that this structure yields a mode volume as low as \unit{0.7}{\micro\meter^3} and that a large coupling 
efficiency of over 0.8 can be achieved (see Methods and Supporting Information)~\cite{Almokhtar2014}. 
The corresponding electric field distribution is shown in Figure~\ref{fig1}c while Figure~\ref{fig1}d shows the 
calculated transmission spectrum. 
For the cavity mode that appears in the middle of the  Bragg gratings' stop band (linewidth 
of about \unit{10}{\nano\meter}) as a narrow band peak, a numerical Q-factor of about 1600 is found.
Measured transmission spectra of an NFBC are shown in Figure~\ref{fig2}b,c. 
Here, due to imperfections probably caused by drifts during FIB fabrication, 
the achieved Q-factors are reduced to a value of about 250.

In order to tune the resonance wavelength of the cavities, mechanical tension along the fibre's axis can be applied 
using a one-axis piezo actuator (see Figure~\ref{fig2}a). 
This allows the resonace to be shifted in a very controlled way, 
%When the fibre is tensioned, the resonance shifts in a very controlled way, 
as shown in Figure~\ref{fig2}c. 
The slope of the tuning curve is \unit{0.05}{\nano\meter\per\micro\meter}, which means that an easily controllable step of 
\unit{1}{\nano\meter} results in a wavelength shift of only \unit{37}{\mega\hertz}.  
Resonance shifts of up to \unit{25.8}{\nano\meter} can be achieved  
before the nanofibre ruptures (see Supporting Information).  
The hysteresis of the resonance on tuning is shown in Figure~\ref{fig2}c. 
The resonance shows linear and reversible behaviour in the wavelength range of over \unit{15}{\nano\meter}. 
Notably, transmittance and cavity Q-factor are almost 
constant when tuning the NFBC (variance in Q-factor and transmittance are 4\,\% and 0.7\,\%, respectively).
As shown in the Supporting Information, tuning can also be performed at cryogenic temperatures.

Next, we couple light emitters to our NFBCs.
For this task, random approaches as for example described in reference~\cite{Fujiwara2011} are not 
suitable since the emitter has to be placed exactly at the cavity's position -- in our case a target area just 
\unit{450}{\nano\meter} in length. 
Hence, we developed the following approach:
A sharp tungsten tip
is coated  with particles and brought into contact with the nanofibre. Using an 
optical microscope and an alignment laser coupled to the fibre,
we ensure that the tip only touches the fibre at the cavity's position.  
On retraction of the tip, it is highly probable that only one, or a few, particles 
are deposited. In the following, we show coupling of QDs (emission wavelength \unit{630}{\nano\meter}) 
to the NFBCs.

\begin{figure}
	\includegraphics[width=.7 \textwidth]{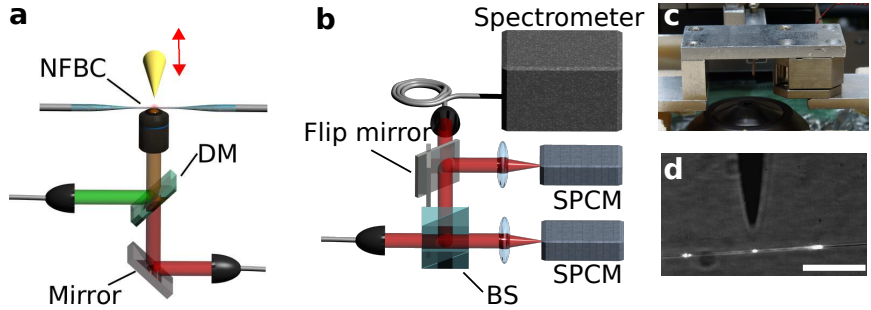}
	\caption{\textbf{Experimental setup.}  
  \textbf{a}, Sketch of the microscope setup. An NFBC is placed in front 
  of a microscope objective, which is used either for fibre coupled confocal microscopy or 
  wide field imaging. A tip sharp is used to place nanoparticles on the fibre in a controlled way. 
  DM is a dichroic mirror.
  \textbf{b}, Sketch of the fibre coupled detection system used.  Two single photon counting modules (SPCM) 
  in Hanbury Brown and Twiss (HBT) configuration are used to measure the photon stream. One arm 
  of the HBT setup can also be coupled into a spectrometer.
  \textbf{c}, Photograph of the NFBC mounted on the setup.
  \textbf{d}, Camera image of the coupling process. Scalebar is \unit{20}{\micro\meter}.
  }
	\label{fig:setup}
\end{figure}

The experimental setup used to investigate coupling of emitters to our 
NFBCs (see Figure \ref{fig:setup}) consists of a microscope 
with a three-axis translation stage and a tensionable fibre mount. 
Light sources (lasers, halogen lamp) and detection units (spectrometer, 
avalanche photodiodes in Hanbury Brown and Twiss configuration) are all fibre coupled making it 
easily possible to
attach the sources/detectors either to the ends of the tapered fibre or to the microscope.
\begin{figure}
	\includegraphics[width=.7 \textwidth]{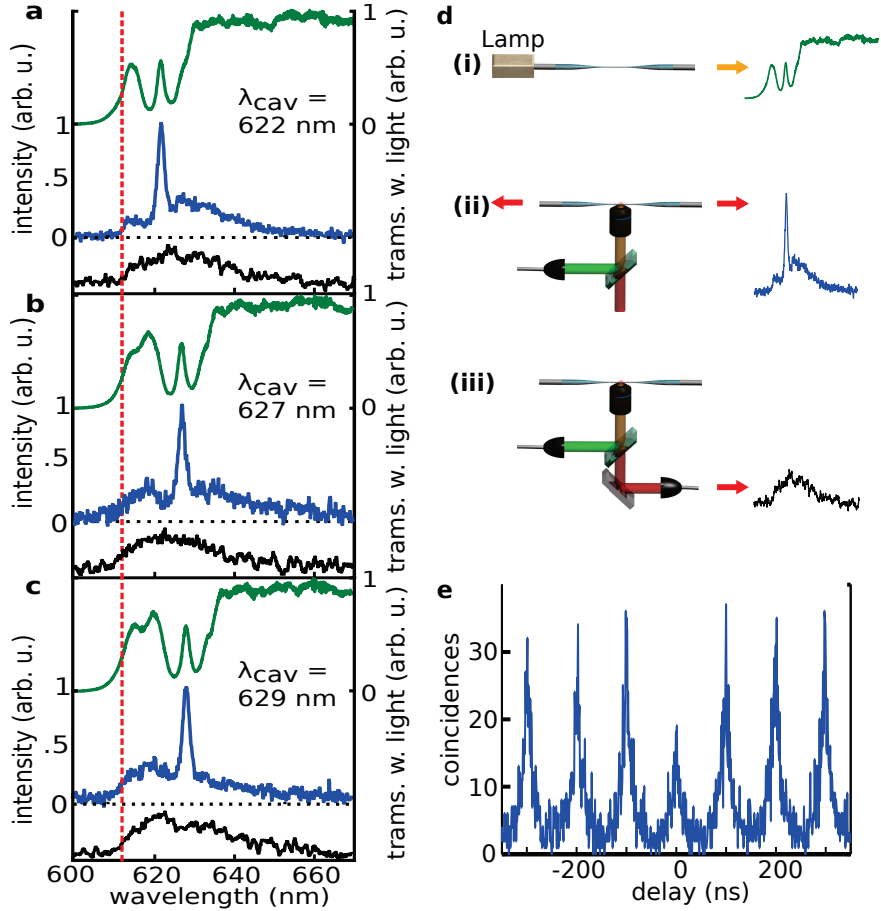}
	\caption{
  \textbf{Fluorescence spectra of a few QDs coupled to a nanofibre cavity.} 
  \textbf{a,b,c},
  Measured signal for three different tunings 
  (\unit{622}{\nano\meter},\unit{627}{\nano\meter}, and \unit{629}{\nano\meter}) of
  the fibre nanocavity. Spectra acquired using a halogen lamp 
  are shown in green (\textbf{i} in panel \textbf{d}), spectra acquired through the fibre while exciting 
  using the microscope objective are shown in blue (\textbf{ii} in panel \textbf{d}), and spectra acquired 
  using the confocal microscope are shown in back (\textbf{iii} in panel \textbf{d}).
  The dashed vertical line indicates the cut-on wavelength of the longpass filters used 
  and the black curves are offset by \unit{-0.2}{arb. u.} for clarity.
  \textbf{d}, Sketch of the different setups used.
  \textbf{e}, Antibunching measurement in confocal 
  configuration under pulsed excitation. From the clear dip going below 0.5 it can be 
  concluded that the main contribution to the signal stems from a single emitter.
  }
	\label{fig:fewspec}
\end{figure}
Figure~\ref{fig:fewspec} shows measurements of a few QDs coupled to an NFBC. 
Three different experimental configurations (see Figure~\ref{fig:fewspec}d) are used. 
In the first, a laser of \unit{532}{\nano\meter} wavelength is focussed via the microscope 
objective and fluorescence is collected using the same objective in a confocal arrangement 
(black curves in Figure~\ref{fig:fewspec}a-c).
In the second configuration, light from a halogen lamp is coupled into the fibre and transmitted light is 
measured (green curves in Figure~\ref{fig:fewspec}a-c). Finally, in the third configuration, the 
emitters are again excited by the laser through the objective lens, and the light coupled into both fibre ends is 
analysed (blue curves in Figure~\ref{fig:fewspec}a-c).
While the confocally detected spectrum exhibits a Gaussian shape, the light detected at the fibre's ends 
clearly shows an enhancement at the cavity's resonance and suppression in the stop band of the Bragg gratings. 
When the NFBC is tuned, the enhanced emission clearly shifts with the cavity resonance. 
Note that in contrast to  temperature tuning 
 or gas condensation tuning via tension leaves the emitter unaffected. 
This is a clear indication that the cavity is the sole source of the fluorescence enhancement. 
An evaluation of the fluorescence maxima detected through the fibre and comparison with the 
confocal measurement yields enhancement factors of approximately 2.7 for all three cavity tunings shown,
which is in excellent agreement with theoretical predictions. More specifically, when 
taking into account the measured Q-factor of this NFBG ($Q=330$) and the mode volume of \unit{0.7}{\micro\meter^3}
derived from the FDTD calculations, an enhancement factor of approximately 3 is 
expected when the Purcell formula is used. 
Using improved structures with Q-factors closer to the calculated value,  
this enhancement could even reach 14. Note that this is a first implementation and 
it is highly probable that significant improvements to this value can be achieved by 
more sophisticated designs and more accurate fabrication, as it 
was done with two-dimensional photonic crystal and nanobeam cavities.

\section*{Conclusion}

In conclusion, we have introduced the system of nanofibre Bragg cavities with small mode volumes on the 
order of a cubic wavelength and an ultra-wide tuning range. The  NFBGs are intrinsically 
fibre coupled, which, together with 
their high quality factors and small mode volumes, makes them ideally suited for use in altering the 
photonic properties of quantum emitters. A hysteresis-free tuning with a range of more than 
\unit{20}{\nano\meter}, which is achieved simply by straining the 
NFBC, enables adjustment of cavity-emitter coupling without changing 
other experimental parameters such as temperature or surface layers. 
The NFBCs can be operated and tuned 
under cryogenic conditions, making it feasible to couple them with emitters that need cryogenic temperatures. 
We have also shown controlled coupling of QDs to NFBCs resulting in an enhancement of their 
emission by a factor of 3 -- an enhancement that probably can be increased by more 
advanced design and fabrication techniques.
Note that, in principle, other particles, e.g. nanodiamonds containing defect colour 
centres~\cite{Liebermeister2014}, can also be coupled to our NFBC  
using the technique introduced here. 
With these properties, NFBCs will improve single photon sources for nanophotonic quantum networks and will enable   
efficient implementation of important devices needed for quantum information science, such as single photon switches.

\section*{Methods}
\textbf{Fabrication of nanofibre cavities.} 
A focused ion beam milling system (SMI-2050, Seiko Instruments Inc.) is used to etch the cavity structures. 
The nanofibres are fabricated by heating a single-mode optical fibre with a ceramic heater and stretching it 
into a fine thread. Ga$^+$ ions of \unit{30}{\kilo\volt} accelerating voltage and 
\unit{9.3}{\pico\ampere} beam current are focused on the 
nanofibre with the spot size of \unit{13}{\nano\meter} and scanned perpendicularly to the direction of the long axis of the 
fibre in order to fabricate a 160 periods grating with one defect in the middle. \\
\textbf{Numerical calculations.} 
In order to analyse the fibre cavities, 3D finite-difference time-domain (FDTD) simulations are  
performed using a commercial package (FDTD Solutions, Lumerical). The modelling 
geometry, which is simplified from the structure utilized in the experiment, can be found in the 
Supporting Information. Grooves with a period of \unit{300}{\nano\meter} 
and a depth of \unit{45}{\nano\meter} are carved on the upside and transverse sides of the nanofibres 
(diameter \unit{300}{\nano\meter}). A \unit{450}{\nano\meter} 
defect is introduced in the centre of the 160 period grating. 
The simulation region is 60 $\times$ 2 $\times$ 2 $\mu$m and perfectly matched layers (PML) are 
employed as absorbing boundary conditions. Time steps and simulation time are 
\unit{0.039}{\femto\second} and \unit{5}{\pico\second}, respectively. 
We use an automatic non-uniform mesh with a high accuracy and the 
 material properties of SiO$_2$ provided by Lumerical in the calculation. The light source is placed at a location 
 \unit{-24.5}{\micro\meter} away from the centre of the simulation region and excited the fundamental mode of 
the nanofibre. The transmittance is 
monitored at  \unit{25}{\micro\meter}  away from the centre of the simulation region.\\
\textbf{Emitter-cavity coupling.} 
For coupling emitters and cavities colloidal quantum dots (QSP630, Ocean Nano Tech) dispersed in 
toulene are used. A sharp tungsten tips (TP-0002, Micro Support) is dipped in the 
QD solution. 
An appropriate concentration of QDs in toluene
(approx. 0.06 mg/ml) is used to ensure that only a few QDs stick to the tip's apex. 
Using a three 
axis translation stage equipped with stepper motors and piezo actuators, the 
tip's apex is brought in contact with the cavity and retracted subsequently. 
The process is monitored using the microscope described below and an off-resonant 
laser is coupled to the fibre to provide additional optical feedback via scattering of the 
evanescent field due to the tip. \\
\textbf{Optical measurements.}
The microscope used is an IX 71 (Olympus) with an high numerical aperture objective (MPLAPON 100$\times$/0.95). 
Optical images are acquired using a CCD camera (PRO EM 512 B, Princeton instruments). During confocal operation, a laser beam is sent  
through the back port and the emitted fluorescence light is coupled in 
a multimode fibre (P1-1550A-FC-2, Thorlabs) using a fibre coupler intalled after the tube lens. 
Spectra are acquired using a spectrometer (MS257, ORIEL Instruments) with a CCD camera (DU420-OE, Andor). 

Spectra in Figure~\ref{fig2} are normalized to the transmission of 
a non-tapered single mode fibre and the blue curves in 
Figure~\ref{fig:fewspec} are averages of both fibre ends.
The avalanche photodiodes used in the HBT are SPCM (Perkin Elmer). \\

\section*{Acknowledgement}
We would like to thank to Dr. Oshima at ISIR for his technical support in the early stage of our FIB use. 
We gratefully acknowledge financial support from MEXT-KAKENHI Quantum
Cybernetics (No.
21101007), JSPS-KAKENHI (Nos. 26220712, 23244079, 25620001, 23740228, 26706007,
and 26610077), JST-CREST, JSPS-FIRST, the Project for Developing
Innovation Systems of
MEXT, the G-COE Program, and the Research Foundation for Opto-Science
and Technology.
MF extends thanks to the fellowship from Alexander von Humboldt Foundation and
Yamada Science Foundation. This project was supported by JSPS and DPG under the Japan -- Germany Research Cooperative Program.
A portion of this work was supported by "Nanotechnology Platform Project (Nanotechnology Open Facilities in Osaka University)" 
of MEXT, Japan [F-13-OS-0017, F-14-OS-0004].

\section{Author contributions}
AWS, HT, SK, YO, and MF performed the experiments. Numerical
calculations were carried out by HT. OB and ST supervised the study.
AWS, HT, and ST wrote the paper and all authors discussed the experiment
and text.

\end{document}

\section*{Nanofibre Bragg cavities}
\begin{figure}[h]
	\includegraphics[width=.7\textwidth]{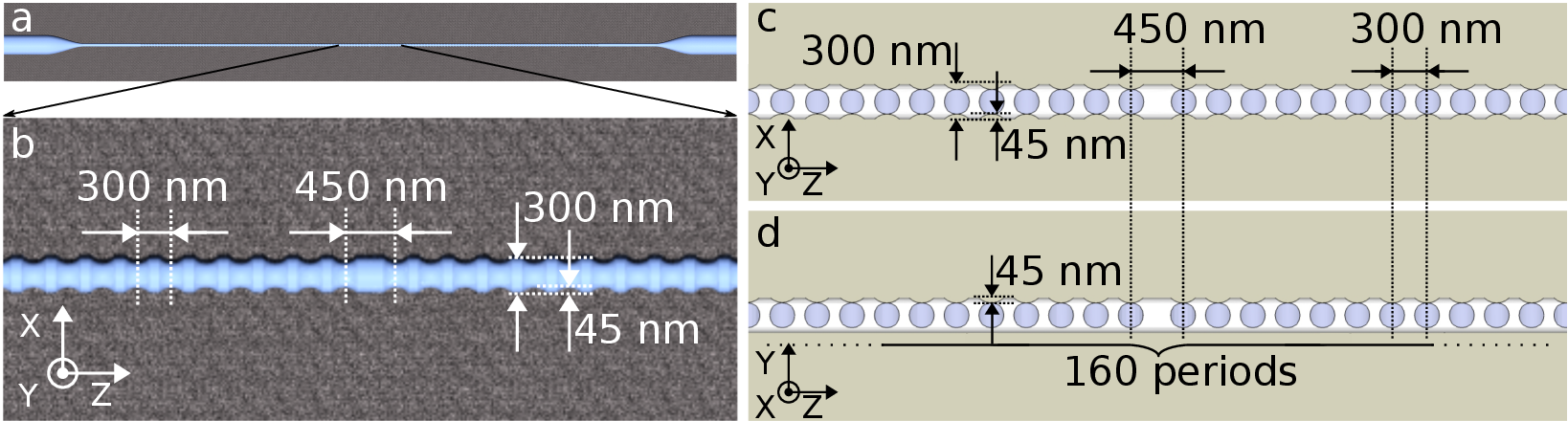}
	\caption{Nanofibre Bragg cavities. \textbf{a} and \textbf{b} 
  are schematic views of the NFBCs while \textbf{c} and \textbf{d} 
  show simplified structures as used in the numerical calculations}
	\label{s1}
\end{figure}

Nanofibre Bragg cavities (NFBCs) are
created by tapering down a 
standard optical fibre (630 HP, Thorlabs) using a homebuilt tapering machine to a diameter of 
approximately \unit{300}{\nano\meter} and then using a focused ion beam to mill periodic groves into the tapered 
region. The grooves are milled from one side, resulting in an arch-like shape. Their depth is 
\unit{45}{\nano\meter}. To yield a Bragg grating with a design wavelength of \unit{630}{\nano\meter}, 
the spatial period of the grooves $\Lambda_B$ is adjusted according to the Bragg formula:
\begin{equation}
\lambda_B = 2 n_{eff}\Lambda_B,
\end{equation}
where {\it n$_{eff}$} is the effective refractive index of the nanofibre. Putting in the 
calculated effective index {\it n$_{eff}$=1.07} for the wavelength of \unit{630}{\nano\meter} 
and the fibre diameter of 
\unit{300}{\nano\meter}, the grating period $\Lambda_B$ is determined to be \unit{300}{\nano\meter}.
The grooves achieve a large refractive index modulation $\Delta n$ of $\sim$ 0.066 as calculated via 
finite difference time domain simulations. 
In the middle of the Bragg grating, a defect of $\frac{3}{2}\Lambda_B$ is introduced in order to form the 
cavity. 
The resulting structure is shown in Figure~\ref{s1}a,b. Figure~\ref{s1}c,d show the 
simplification of the NFBC structure we used in the numerical calculations.

\section*{Simulation of the coupling efficiency}
\begin{figure}[h]
	\includegraphics[width= .4\textwidth]{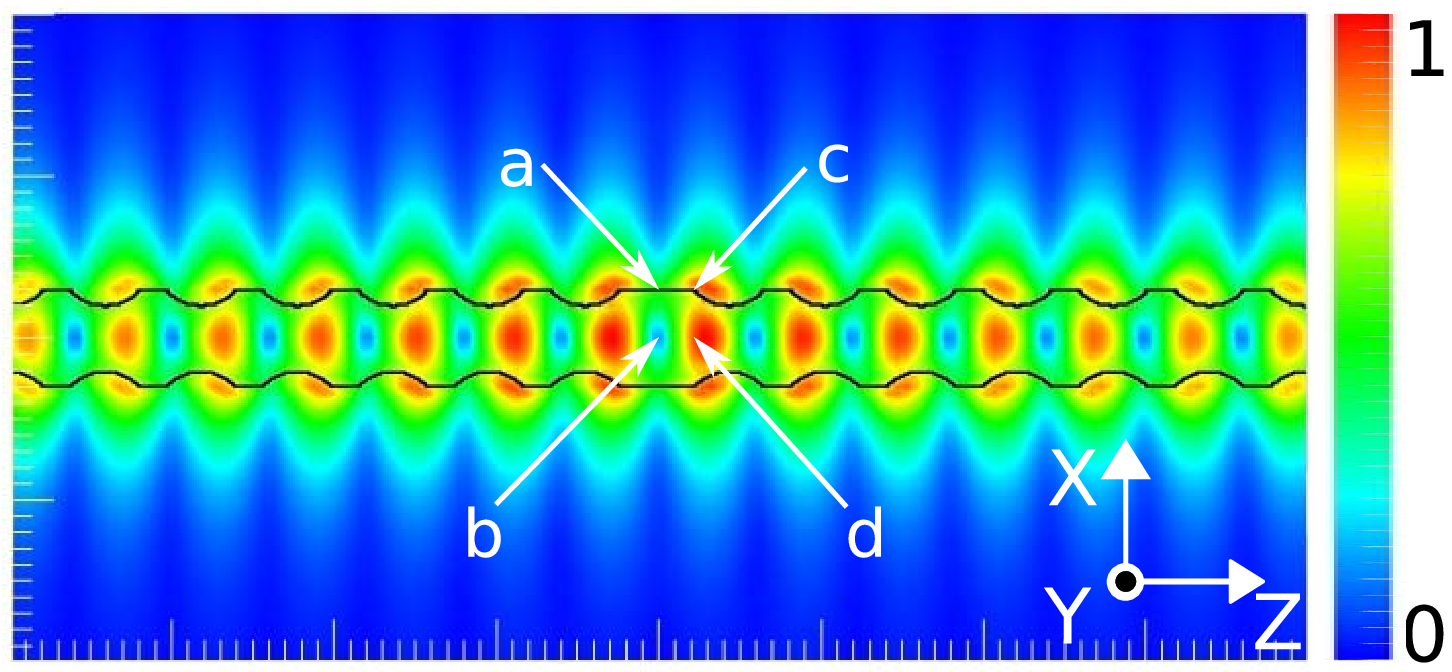}
	\caption{Electric field at a NFBG. A cross-section of the electric field 
  distribution at the centre of the fibre about the Y axis (Y = \unit{0}{\micro\meter}). 
  The black line indicates the cavity structure.
  The positions of the dipoles  
  (a, b, c and d) for the simulation of the 
  used in the simulations for the coupling efficiency are marked by arrows.
  }
	\label{ceff}
\end{figure}

The coupling efficiencies to a fundamental mode of the fibre are calculated based on our 
recent paper~\cite{Almokhtar2014}.
Figure~\ref{ceff} shows the geometry used for the simulation. The dipoles are placed at weak and strong points of 
the electric field inside the cavity. The direction of the dipoles is the X-axis. The wavelength of the 
dipoles is set to the resonance of the NFBC (\unit{634.32}{\nano\meter}). The coupling efficiencies for one end 
of the fibre are monitored at \unit{28}{\micro\meter} away from the centre of the simulation region and 
then summed up for the total coupling efficiency. 

Table 1 shows the calculation results for the cases where a single dipole is located 
at the corresponding position. When the dipole is placed at positions with a weak electric field 
(a and b), the coupling efficiency is very low. In contrast, when the dipole is 
placed at positions with a strong electric field (c and d), the coupling efficiency is over 0.8.
\begin{table}
 \begin{tabular}{c|c|c|c}
&X ($\mu$m) & Z ($\mu$m) & \shortstack{Coupling\\efficiency} \\ \hline
\ a \  & 0.15 & 0 & 0.009  \\ \hline
\ b \ & 0 & 0 & 0.013  \\ \hline
\ c \ & 0.15 & 0.118 & 0.812 \\ \hline
\ d \ & 0 & 0.118  & 0.838 \\ \hline
\end{tabular}
\caption{Calculation results for the coupling efficiencies for four dipolar positions.}
\end{table}

\section*{Tuning at cryogenic temperatures}
\begin{figure}[h]
	\includegraphics[width= \textwidth]{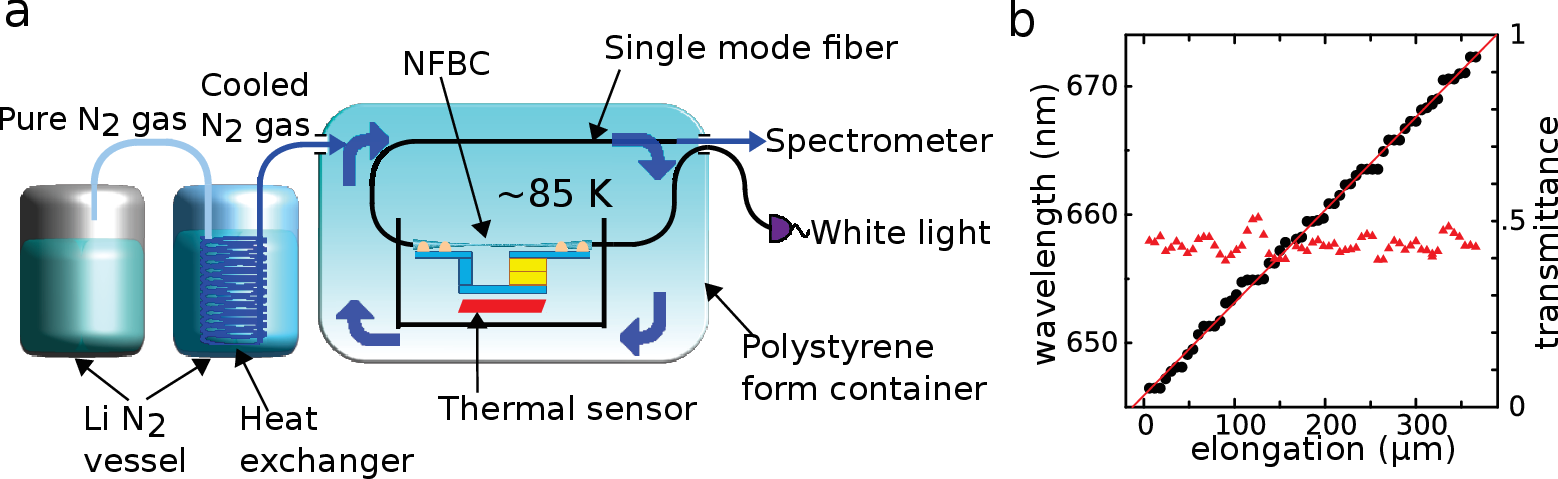}
	\caption{Tuning at cryogenic temperatures. \textbf{a} shows a schematic 
  diagram of our home-made cooling system. Pure nitrogen gas is cooled in a heat exchanger and then 
  used to cool down the NFBG, which can be tuned and is connected to a light source and 
  a spectrometer. \textbf{b} shows a tuning measurement performed at \unit{85}{K}. Black dots
  are the resonance wavelength while the red triangles are the corresponding transmission. The red 
  line is a fit to the data.}
	\label{s3}
\end{figure}

Figure~\ref{s3}a shows our homebuilt cooling system for tuning at cryogenic temperatures. 
The system consists of a volume thermally isolated from the environment by a polystyrene foam container, in which 
the  translation stage hosting the fibre cavity is fixed on an additional metal case. To monitor 
the temperature around the sample, a T type 
thermocouple is fixed next to the stage. For the input and output of light and 
refrigerant, two feed-throughs are mounted on the walls of the container. Single 
mode optical fibres are directly installed in one of the container feedthroughs
and connected to the fibre cavity using a fibre splicer. To prevent icing 
on the fibre, evaporated nitrogen gas is employed as the refrigerant. Pure 
nitrogen gas is cooled with a heat exchanger and flowed through the container. 
Temperatures reached as low as \unit{85}{K} in approximately 30 minutes. 

Figure~\ref{s3}b shows the resonance shift on tensioning the fibre and the corresponding transmittance of a NFBC at \unit{85}{K}.
 Its resonance wavelength is red-shifted linearly in a range of \unit{25.8}{\nano\meter}. 
The transmittance is almost constant during the tuning. No significant transmittance reduction 
is observed between the room temperature and \unit{85}{K} (not shown).  

\section*{Coupling of QDs to tapered fibre cavity}
\begin{figure}[h]
  \centering
	\includegraphics[width= .8\textwidth]{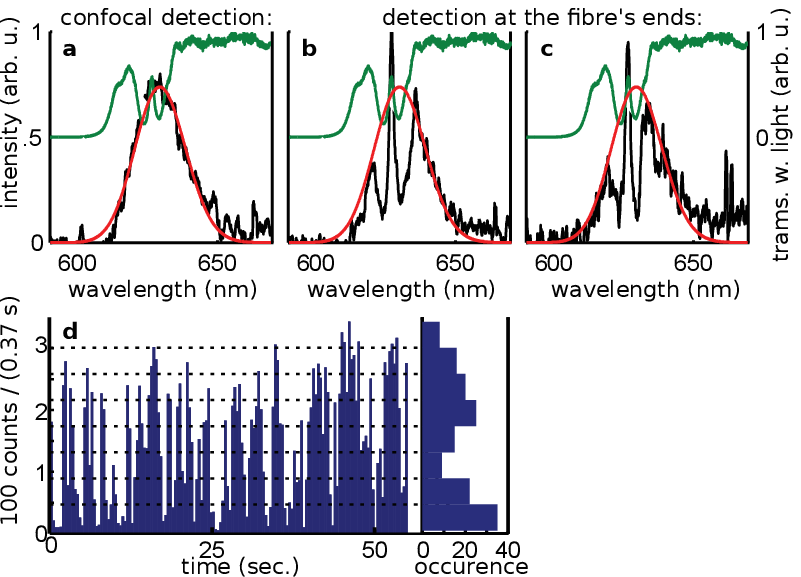} 
	\caption{Fluorescence spectra of a single quantum dots (QDs) coupled to a nanofibre cavity. 
  Panels \textbf{a,b,c},
  show the detected spectra in confocal configuration and at the fibre's ends. 
  The measured data is shown in black and a
  a Gaussian energy distribution fitted to the confocal measurementsis shown in red. In green the corresponding
  cavity resonance is shown. An enhancement of the single QD's fluorescence is visible at the resonance wavelength 
  as well as strong suppression in the Bragg mirrors' band gap. This behaviour indicated that the QD's 
  spatial position is not exactly at the cavity, but slightly shifted towards one of the mirrors.
  In \textbf{d} a time trace of the QD's emission as collected through the fibre is shown. 
  A clear two-level blinking indicates the presence of a single quantum emitter.   
  }
	\label{fig:onespec}
\end{figure}
In order to get to the single-emitter regime -- the regime needed, for example, for efficient 
single photon sources -- we bleach the emission from the quantum dots with the excitation laser until 
the emission is considerably darker. 
The resulting emission spectrum is shown in 
Figure~\ref{fig:onespec}. While the main features are similar to the case of a few QDs, the enhancement 
is less pronounced while the emission inhibition inside the band gap is larger. This means that in 
the bleaching process a QD survived that is not perfectly coupled. On the other hand, this also means 
that coupling for the now bleached particles was even better than calculated. To prove the single emitter 
character of the remaining quantum dot, we look at its blinking behaviour as shown in Figure~\ref{fig:onespec}d 
where two-level blinking is visible -- a clear indication for single quantum systems~\cite{Empedocles1999}.

\bibliographystyle{unsrt}
\bibliography{fibrecavitiy}